\documentstyle[twoside,fleqn,mssl_workshop,psfig]{article}
\newcommand{\etal}{{\it et al.} }

\newcommand{\rosat}{{\it ROSAT} }
\newcommand{\xmm}{{\it XMM-Newton} }
\newcommand{\chandra}{{\it Chandra} }

\begin{document}

\title{Multi-wavelength observing of a forming solar-like star}

\author{S. G. Gregory
\address{SUPA, School of Physics and Astronomy, Univ. of St Andrews, St Andrews, KY16 9SS, U.K.}
E. Flaccomio 
\address{INAF - Osservatorio Astronomico di Palermo, Piazza del Parlamento 1, 90134 Palermo, Italy}
C. Argiroffi$^{\,{\rm b}}$
J. Bouvier
\address{Lab. d'Astrophysique, Univ. J. Fourier, CNRS, UMR 5571, BP 53, 38041 Grenoble, France}
J.-F. Donati
\address{LATT - CNRS/Univ. de Toulouse, 14 Av. E. Belin, F-31400 Toulouse, France}
E. D. Feigelson
\address{Dept. of Astronomy and Astrophysics, 525 Davey Lab., Pennsylvania State Univ., University Park, PA 16802, U.S.A.}
K. V. Getman$^{\,{\rm e}}$
G. A. J. Hussain
\address{ESO, Karl-Schwarzschild-Str. 2, Garching bei Mu\"{n}chen, D-85748, Germany}
M. Jardine$^{\,{\rm a}}$
F. M. Walter
\address{Dept. of Physics and Astronomy, Stony Brook Univ., Stony Brook, NY 11794-3800, U.S.A.}
}

\begin{abstract} 
V2129~Oph is a $1.35\,{\rm M}_{\odot}$ classical T Tauri star, known to possess a strong and complex magnetic
field. By extrapolating from an observationally derived magnetic surface map, obtained through Zeeman-Doppler
imaging, models of V2129~Oph's corona have been constructed, and used to make predictions regarding the global X-ray
emission measure, the amount of modulation of X-ray emission, and the density of accretion shocks. In late June 2009
we will under take an ambitious multi-wavelength, multi-observing site, and near contemporaneous campaign,
combining spectroscopic optical, nIR, UV, X-ray, spectropolarimetric and photometric monitoring. This will allow the
validity of the 3D field topologies derived via field extrapolation to be determined.
\end{abstract}

\maketitle

%------------------------------------------------------------

\section{Introduction}
Classical T Tauri stars (CTTS) represent a key transitional period in the life of a star,
between the embedded protostellar phase of spherical accretion and the main sequence stage. They
are low mass pre-main sequence stars which accrete material from dusty circumstellar disks. They possess 
strong magnetic fields, of order a few kG \cite{joh07}, which truncate the disk and force in-falling gas to flow along
the field lines. Material rains down on to the stellar surface, where it shocks and produces hotspots that emit 
in the optical, UV, and X-ray wavebands.  

CTTS can be in excess of 1000 times more active in X-rays than the Sun is presently. X-rays from the central
star may influence the dynamics and chemistry of the circumstellar disk, which will in turn set the initial conditions for planet formation, 
e.g. \cite{fei02,pas07}.  Understanding the distribution of the emitting plasma
is important to assess the magnitude of these effects. The detection of rotationally modulated X-ray 
emission suggests that the bulk of the emitting plasma is confined within compact magnetic
loops \cite{fla05}, but emission may also arise from large scale loops extending to several stellar
radii \cite{get08}.  Excess soft X-ray emission is another characteristic of CTTS \cite{tel07}.  However, in only 5 systems is this soft excess
observationally associated with the dense plasma in accretion shocks (see \cite{gue09} for a review).  
As a further indication of the complexity of CTTS, X-ray emission may also arise from shocks in outflowing bipolar jets
\cite{gue07}.    
 
In order to model X-ray emission from CTTS we require information about the magnetic fields that 
confine the coronal plasma and drive X-ray flaring via reconnection events.  In this article 
we describe how models of the coronae of CTTS can be constructed via field extrapolation from surface
magnetograms derived through Zeeman-Doppler imaging.  We describe how a large multi-wavelength, multi-observing site campaign on the 
CTTS V2129~Oph will be used to test the ability of such models to capture the true 3D structure
of CTTS magnetospheres, and their ability to predict X-ray emission properties.  

%---------------------------------------------------------

\section{ZDI and field extrapolation}
\begin{figure}[!ht] % fig.1
\vspace{10pt}
\centerline{\psfig{file=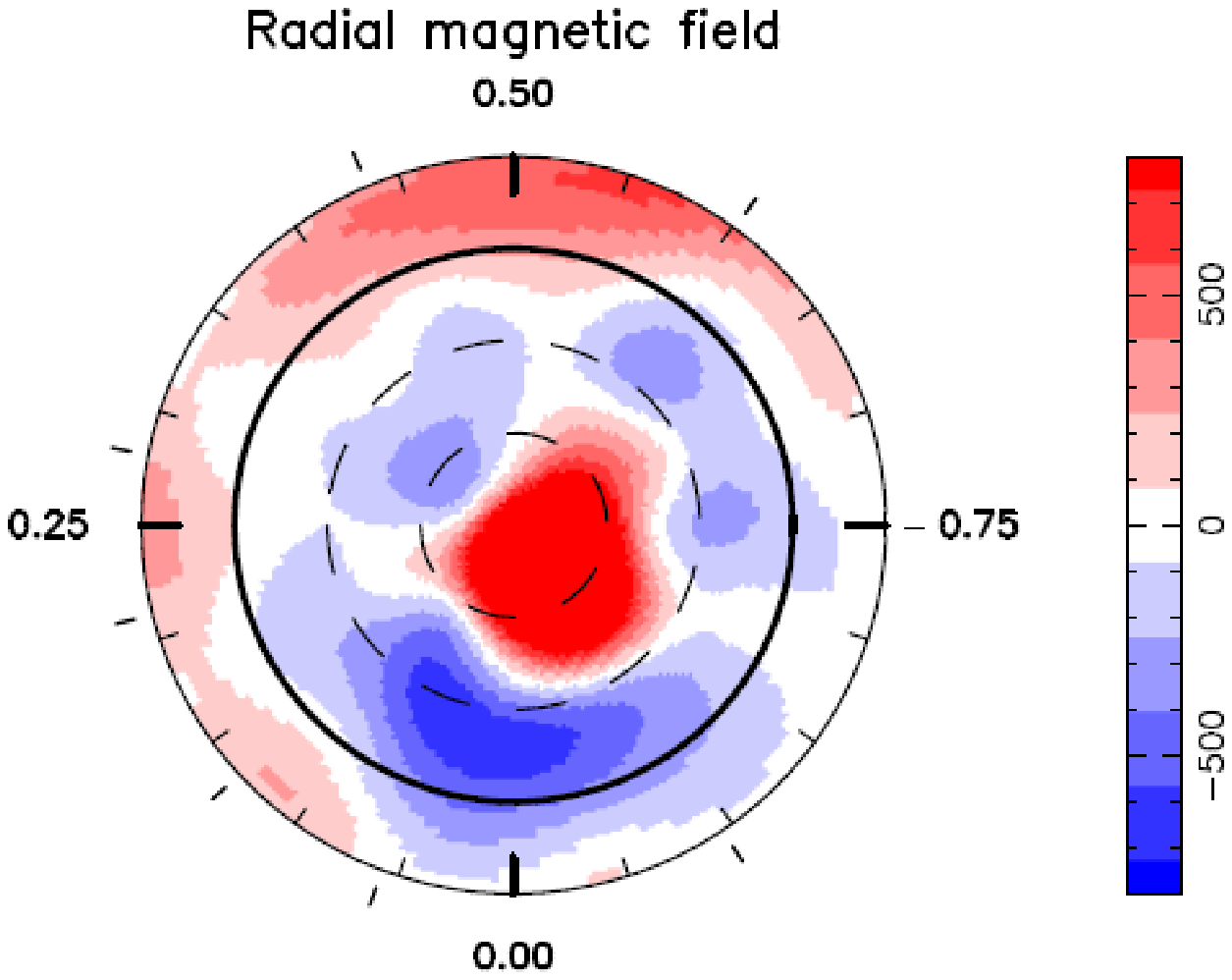,width=5.875cm}}
\vspace{10pt}
\centerline{\psfig{file=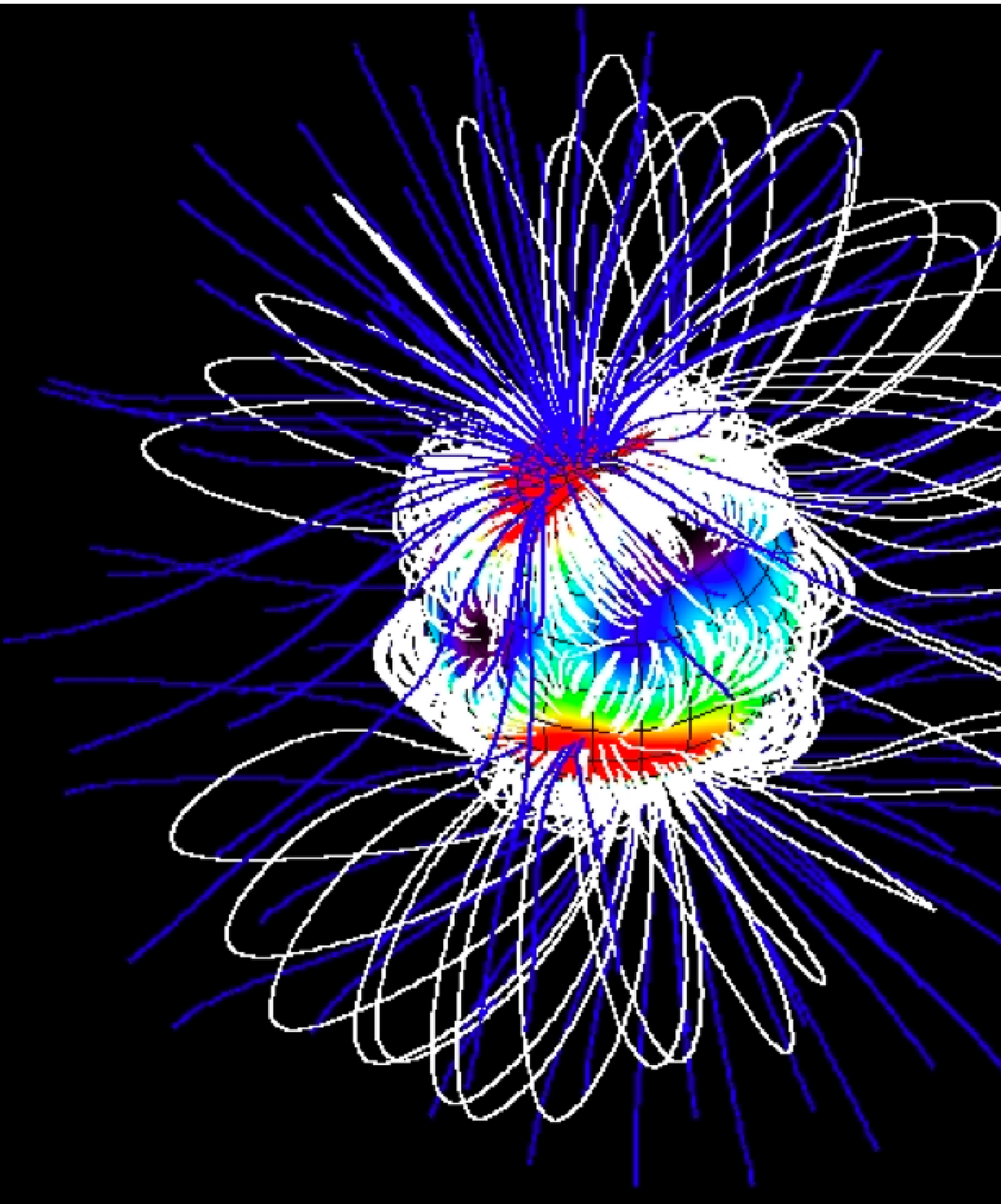,width=4.5cm}}
\caption{
{\it Top panel}: A flattened polar projection of the surface magnetic topology of the accreting T Tauri star V2129~Oph,
 with the equator depicted as a bold circle \cite{don07}.  {\it Bottom panel}: Red (blue) corresponds to positive (negative) 
field polarity with the fluxes labeled in Gauss.  A field extrapolation showing the simple large scale field which 
interacts with the circumstellar disk, and the complex surface field which contains X-ray emitting plasma.  Open 
field lines are blue, with closed field lines in white.
}
\label{extrap}
\end{figure}

Using the technique of Zeeman-Doppler imaging (ZDI) it is possible to map the
medium and large scale structure of stellar magnetic fields (see \cite{don09} for a review of the technique).  For CTTS, such surface maps are constructed
by monitoring the rotational modulation of the Zeeman signal in both photospheric absorption lines, and 
emission lines which form at the accretion shock.
Fig. \ref{extrap} shows an example magnetic surface map of the CTTS V2129~Oph.  The magnetic field of this
star was found to be dominantly octupolar, with a contribution from many higher order field components \cite{don07}.

By assuming a potential field the 3D structure of CTTS magnetospheres can be determined via field extrapolation    
from observationally derived magnetic surface maps \cite{gre08}.  Fig. \ref{extrap} shows such an extrapolation 
from the magnetic map of V2129~Oph \cite{don07,jar08}.  A clear distinction can be made between the complex surface
field regions, with contains the $\sim10^7\,{\rm K}$ plasma that constitutes the X-ray emitting corona, and the 
somewhat simpler larger scale field that interacts with the circumstellar disk.  The larger scale field, however, is 
distorted close to the stellar surface by the strong and complex surface field regions \cite{gre08}.

%---------------------------------------------------------

\section{Modeling the stellar X-ray emission}
The bulk of the X-ray emission from CTTS is thought to arise from magnetic reconnection events (flares)
in a scaled-up solar-like coronae \cite{pre05}.  Once the structure  of the surface magnetic field has been determined
via field extrapolation, models of the distribution of X-ray bright regions can be constructed \cite{gre06}.       
If we assume that the coronal plasma along each field line loop is in hydrostatic equilibrium then the gas pressure is
given by,
\begin{equation}
p(s) = p_0 \exp{\left( \frac{\mu m_H}{k_B}\int_s \frac{\mathbf{g}\cdot\mathbf{B}}{BT}ds \right )},
\end{equation}
where $p_0$ is the base pressure at the stellar surface (at the loop footpoints), $s$ is the coordinate along the loop, and the field vector $\mathbf{B}$
is calculated at each point within the stellar magnetosphere by field extrapolation \cite{jar08}.  The coronal plasma can either be assumed 
to be isothermal, or the temperature along individual loops scaled according to the loop length and base pressure.  At each point along the field
line loops we calculate the plasma-$\beta$, the ratio of gas to magnetic pressure.  If $\beta > 1$ at some point along a field line then the loop
is assumed to be unable to contain the coronal plasma, and is distorted and blown open, and is therefore dark in X-rays.  Thus there is an 
inhomogeneous distribution of X-ray bright regions across the stellar surface.  The X-ray dark areas contain the 
footpoints of wind bearing open field lines, and those of the larger scale field lines which interact with the disk.  The latter regions, however,
may also contain accretion shocks that emit at softer X-rays energies (see below).  Once the pressure
(or equivalently the gas number density $n_e$) is known at each point within the corona the global X-ray emission measure (EM) can be obtained,  
\begin{equation}
EM = \int\int\int n_e^2 dV = \frac{1}{4k_B^2}\int\int\int \frac{p^2}{T^2} dV.   
\end{equation}

From the extrapolated magnetic field of V2129~Oph, assuming an isothermal corona at $30\,{\rm MK}$, we calculate an X-ray EM of ${\rm log}EM = 53.9\,{\rm cm}^{-3}$.  
Although this is large for a prototypical $0.5\,{\rm M}_{\odot}$ T Tauri star, it is reasonable
for a more massive star like V2129~Oph (see \cite{pre05}), and agrees within a factor of three with the X-ray luminosity calculated from \rosat
data \cite{cas95}.

\begin{figure}[t] % fig.2
\vspace{10pt}
\centerline{\psfig{file=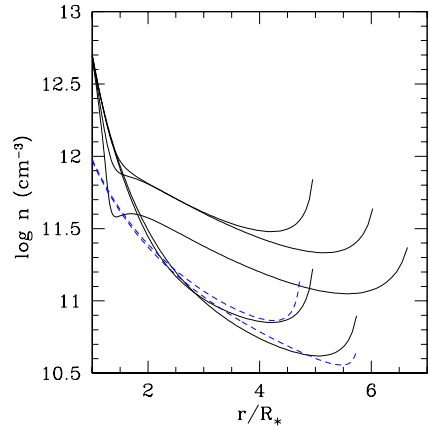,width=2.3in}}
\caption{A steady state accretion flow model suggests that X-ray emission from accretion 
hotspots at the surface of V2129~Oph will be detectable \cite{gre07,jar08}.  The solid black lines represent the 
density structure of infalling gas flowing along the extrapolated field lines having left the inner disk from a range of radii.
The dashed blue lines show the expected gas number density for accretion along dipolar field lines.
}
\label{denpro}
\end{figure}

In addition to coronal X-ray emission, the cool but dense plasma at accretion shocks emit softer X-rays, e.g. \cite{tel07}.  
By assuming that accretion occurs at a steady rate along the larger scale field lines, and that mass and magnetic flux are 
conserved along each accreting flux tube, the number density of the accreting gas along the field lines connecting the disk 
to a particular hotspot can be determined from,
\begin{equation}
n(r) = \frac{1}{\mu m_H}\frac{B(r)}{B(R_{\ast})}\frac{\dot{M}}{A_{\ast}v(r)},
\label{den}
\end{equation}
where $\dot{M}$ is the fraction of the total disk accretion rate in to the particular hotspot that covers
an area of the stellar surface of $A_{\ast}$ \cite{gre07,jar08}.  
Fig. \ref{denpro} show some example density profiles along a selection of accreting field lines assuming 
that accretion occurs from a range of radii within the disk of V2129~Oph.  The shock densities
derived from the model are large enough to expect some soft X-ray emission from the accretion hotspots.  

\begin{figure*}[!ht] % fig.3
\vspace{10pt}
{\psfig{file=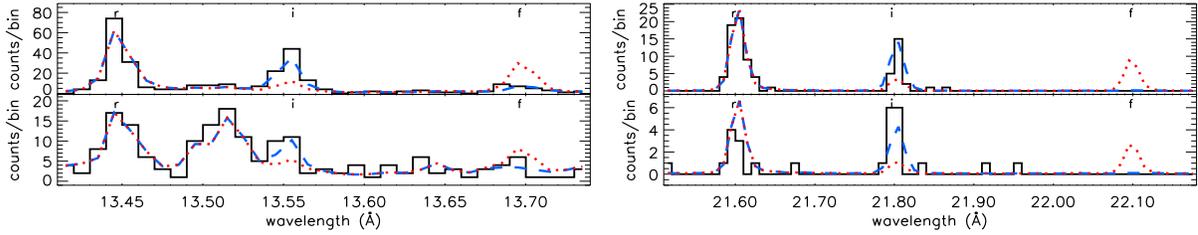,width=6.4in}}
\caption{A simulation of the density sensitive Ne~IX ({\it left panel}) and O~VII ({\it right panel}) line triplets that form in the accretion shock for the 
cool/hot plasma model [{\it upper/lower panels}] assuming ${\rm log}n_e = 13\,{\rm cm}^{-3}$.  The dashed (dotted)
line indicates the noise-free spectra predicted for ${\rm log}n_e = 13 (10)\,{\rm cm}^{-3}$.  
}
\label{spectrum}
\end{figure*}

The few CTTS observed with high resolution X-ray spectroscopy, however, have shown very different average plasma 
temperatures.  In order to understand what we may expect to observe we have simulated \chandra HETG spectra by considering two ``extreme'' 
cases: TW~Hya, which has a cool corona ($T = 2.9, 8.0, 15.5\,{\rm MK}$ and $EM = 19.2, 1.2, 3.1 \times 10^{52}\,{\rm cm}^{-3}$), 
and the hotter BP~Tau ($T = 2.3, 7.3, 25.2\,{\rm MK}$ and $EM = 3.5 ,5.3, 10.3 \times 10^{52}\,{\rm cm}^{-3}$) as determined by \cite{rob06}.  
The models are scaled to match the observed V2129~Oph X-ray flux determined from \rosat data.
Enlargements of the simulated HETG spectra for the hot and cool plasma models in the Ne IX and O VII triplet regions, assuming 
two different plasma densities, and an exposure time of 200 ks, are shown in Fig. \ref{spectrum}. The hot model yields only $\sim$30 counts 
in the O VII triplet, but even in this case, from the expected number of counts in the lines and assuming 
Poissonian statistics, we would be able to distinguish with high confidence between typical coronal 
and accretion shock densities. For plasma densities larger than $\sim 10^{12}\,{\rm cm}$, for which the O VII indicator saturates, the Ne IX 
triplet can be used for the estimation.  In the simulations we have ignored the X-ray background, negligible with respect to our 
expected source count rate. 

%-----------------------------------------------------------

\section{Conclusions: testing the model with multi-wavelength observations}
V2129~Oph has never been observed by \chandra or \xmm.  Our simulated accretion shock density 
based on a flow model using the magnetic field derived via field extrapolation
indicates that X-rays from cool plasma at hotspots should be detectable, and distinguishable from the coronal contribution 
to the X-ray spectra.  The \chandra data, to be obtained in late June 2009 ($2 \times 100\,{\rm ks}$ exposures 
separated by about half a stellar rotation period), will therefore allow the predictions of the
field extrapolation model to be tested.  The surface magnetogram of V2129~Oph, however, was derived from 
spectropolarimetric data obtained with ESPaDOnS at CFHT in 2005 \cite{don07}.  In order to account for possible 
variations in the magnetic field geometry of the star, we are scheduled to obtain new contemporaneous ESPaDOnS observations of 
V2129~Oph, with additional data from the twin instrument NARVAL at TBL in the French Pyr{\'e}n{\'e}es.  A updated magnetic
surface map will be derived using ZDI and new model predictions produced for direct comparison with the X-ray data.

The density of the plasma at the accretion shock determined from the model also depends on the assumed mass accretion rate,
see equ. \ref{den}.  Thus we will obtain additional optical and UV spectra from the HARPS instrument in the southern hemisphere, 
the KPNO 4m telescope and SMARTS (which will also provide photometric monitoring, allowing the distribution of hot/cool spots
at the stellar surface to be inferred) in order to derive an accretion rate through veiling estimates.  Furthermore, nIR spectra 
from CRIRES at VLT will allow Zeeman broadening measurements       
to better constrain the field strength at the surface of V2129~Oph, and will compliment the magnetic geometry determinations from
the spectropolarimetric observations.  The abundance of observational data will allow variability in the accretion related emission lines 
due to rotational modulation, and that due to time variable accretion, to be disentangled.   
The multi-wavelength, near contemporaneous data, in some cases covering several stellar rotation periods, will provide a crucial test of 
the ability of field extrapolation models to reproduce the true 3D nature of the magnetospheres of forming solar-like stars.

%------------------------------------------------------------

%do not change this

%do not change this
\normalsize
                                            
\end{document}